\documentclass[twocolumn]{elsart}
\usepackage{natbib}
\usepackage{graphicx}
\usepackage{amsmath,amssymb}
\def\sun{\odot}

\begin{document}
\runauthor{Li, Wang, Waxman and M\'esz\'aros}

\begin{frontmatter}
\title{Photon Acceleration at Shock Breakout of Trans-Relativistic Supernova}
\author[Weizmann]{Zhuo Li}
\author[psu,nju]{Xiang-Yu Wang}
\author[Weizmann]{Eli Waxman}
\author[psu]{Peter M\'esz\'aros}
\address[Weizmann]{Physics Faculty, Weizmann Institute of Science, Israel}
\address[psu]{Department of Astronomy and Astrophysics, Pennsylvania State University, USA}
\address[nju]{Department of Astronomy, Nanjing University, China}

\begin{abstract}
The predicted thermal flash from supernova (SN) shock breakout might have been detected
for the first time by Swift in GRB 060218/SN 2006aj. The detected thermal X-ray emission
in this event implies emergence of a trans-relativistic (TR) SN shock with kinetic energy
of $E_k\gtrsim10^{49}$erg. During TRSN shock breakout, the thermal photons could be
``accelerated'' by the shock through repeated bulk Compton scattering, forming a
nonthermal $\gamma$/X-ray component with dominant energy over thermal one. This mechanism
of ``photon acceleration'' at TRSN shock breakout might also account for gamma-rays in
the other similar low-luminosity GRBs, implying that they are atypical GRBs with only TR
outflows. TRSNe form a peculiar type of SNe with large kinetic energy,
$\gtrsim10^{49}$erg, in TR ejecta, $\Gamma\beta\gtrsim2$.
\end{abstract}
\begin{keyword}
supernovae, shock waves, gamma-ray
\end{keyword}

\end{frontmatter}


\section{Thermal emission in GRB 060218/SN 2006aj} A SN explosion must result in a shock
wave propagating inside the progenitor star and leading to ejection of the stellar
envelope. The SN shock must be radiation-pressure dominated due to the huge optical depth
trapping the radiation inside the star, and the shock should be mediated by Compton
scattering of photons. The scattering optical depth within the shock width is
$\tau_{sh}\sim c/v_{sh}$ with $v_{sh}$ the shock velocity. As the shock makes its way
outward, the pre-shock optical depth, $\tau$, decreases. Once $\tau\sim\tau_{sh}$, the
radiation start to diffuse faster than shock propagation and hence escapes. At this
moment a thermal radiation flash arises, marking the SN emergence. Shock breakout flashes
have been predicted for decades \cite{history} but never been detected due to their
transient nature and early occurrence in X-ray band.

Recently, thanks to Swift a thermal X-ray component was detected in the prompt emission
of GRB 060218 \cite{obs}. The thermal emission showed total energy $E\sim10^{49}$erg,
temperature $T\approx0.2$~keV and duration $\Delta t\sim10^3$s. For a stellar-scale
event, this amount of energy in keV temperature simply suggests a radiation-dominated
emission region. Its size can be estimated to be $R\sim(E/aT^4)^{1/3}\sim10^{13}$cm, much
larger than that of the compact progenitor for the observed associated type Ic SN. In the
context of SN shock breakout, $R$ should be the breakout radius, so there should be still
substantial material outside of the stellar surface that would trap the radiation. The
required amount of mass beyond the breakout radius $R$ is only $M=4\pi
R^2\tau/\kappa\sim10^{-6}\tau M_\sun$, which could be either a somewhat large stellar
wind (of mass loss rate $\dot{M}=Mv_{wind}/R\sim10^{-4}\tau M_\sun$/yr and wind velocity
$v_{wind}=10^3$km/s) or just some mass shell that is ejected few days before the burst.
Note, $\tau\sim c/v_{sh}$ in the above equations for optical depth at breakout radius
$R$. The shock velocity can be derived from jump condition in the radiation dominated
shock, $aT^4\sim3\rho v_{sh}^2$, where $\rho$ is the medium density at $R$
($\rho=\dot{M}/4\pi R^2v_{wind}$ for a $\rho\propto r^{-2}$ wind). It turns out to be a
TR shock velocity, $v_{sh}/c\sim$ few (more detailed calculation in \cite{Waxman07} gives
$\Gamma\beta_{sh}\sim2$). In short, the observed large thermal energy $E$ with high
temperature $T$ in GRB 060218 might imply that {\em a TRSN shock breaks through from an
optically thick stellar wind or pre-ejected shell.} Note, $R/c\sim200{\rm s}<\Delta t$
just means that this is an anisotropic explosion, where the flash duration is not simply
$R/c$.

\begin{figure*}
\centering
\includegraphics[width=1.1\columnwidth]{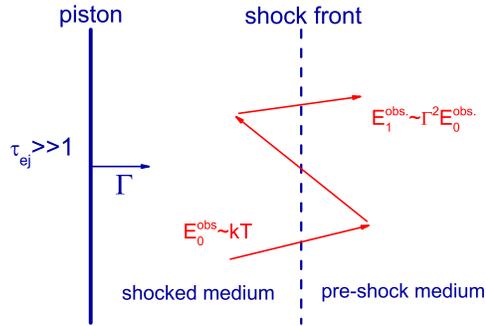}
\caption{Schematic plot of ``shock acceleration" of photons at TRSN shock breakout. The
observed photon energy is enhanced by a factor of $\sim\Gamma^2$ each shock crossing
cycle.}
\end{figure*}

\section{Nonthermal flash: photon acceleration in trans-relativistic SN shock} The
observed prompt emission in GRB 060218 is still dominated by nonthermal $\gamma$/X-rays,
which constitute 80\% flux in Swift XRT. In sense of dominant energy, the production of
nonthermal component is still an important question. We showed in \cite{ph-acc} that the
nonthermal emission is naturally expected in the context of TRSN shock breakout. The
reason is the following. During the TR shock breakout, the pre-shock optical depth
$\tau\sim 1/\beta_{sh}\sim1$ is still substantial, thus some fraction of the outgoing
photons could be backscattered (with a probability of $\sim\tau$) and travel through the
shock from upstream to downstream. These photons would encounter the downstream
electrons, i.e., in the swept-up material or the SN ejecta, which are moving in a bulk
Lorentz factor $\Gamma\sim$few (Downstream electrons are practically cold since their
thermal energy is much less than their bulk kinetic energy). The downstream electrons can
Compton scatter photons back through the shock (from downstream to upstream) with energy
increased by $\sim\Gamma^2$. Then photons might undergo multiple shock crossings, and the
photon energy would be enhanced by a factor $\sim\Gamma^2$ each shock crossing cycle,
leading to a power-law like photon spectrum. This process of ``photon acceleration" (Fig.
1) mimics the Fermi acceleration of charged particles in collisionless shocks. The energy
source for ``photon acceleration" would be the bulk kinetic energy of the SN ejecta, thus
the process can be called bulk Comptonization.

Since the shock propagation leads to that $\tau$ and hence the rate of photon
backscattering are decreasing with time, it is not easy to solve the problem
analytically. In order to find out if photons can be accelerated efficiently and a
dominant nonthermal component arises at TRSN shock breakout, we carry out a monte-carlo
(MC) calculation in this problem with some simple assumptions. We consider the SN ejecta
as a piston with constant $\Gamma$ and infinite optical depth moving in a stratified
medium, $\rho\propto r^{-2}$. The piston drives a shock into the medium, thus there are
three regions in the picture: the piston, the shocked medium and the pre-shock medium.
The thermal photons are generated downstream before breakout. During photons diffuse out
they undergo scattering in these three regions, with energy gain in head-on collisions
and energy lose in head-tail ones. In the MC calculation, we inject thermal photons in
the shocked medium region at some point and follow the scattering history until it
escapes out. Klein-Nishina effect is considered in scattering, and photons are taken as
test particles. Two important assumptions are: first, we consider planar geometry and
photons travelling in one dimension; secondly, we assume infinitesimal shock width,
simply because the structure of radiation-dominated shock is not known for relativistic
case.

\begin{figure*}
\centering
\includegraphics[width=1.1\columnwidth]{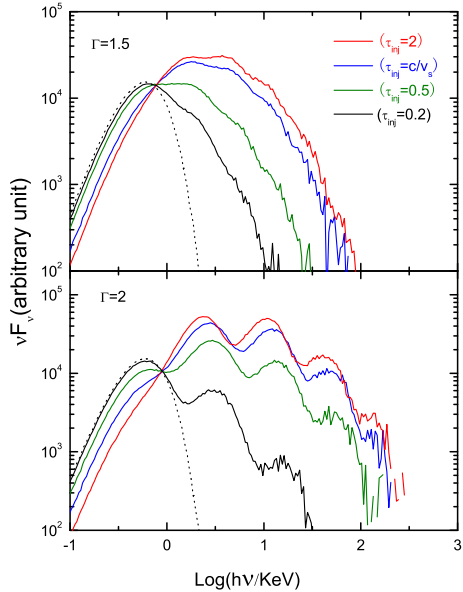}
\caption{MC results: the time-integrated energy distribution of the escaping photons
(solid lines) for single-time injection of thermal photons (dotted lines).}
\end{figure*}

The resulted spectra of calculation for single-time injection of thermal photons,
corresponding to certain $\tau$ or $r$, are shown in Fig. 2 (from Fig. 1 of reference
\cite{ph-acc}). It can be seen that in the case of TR shocks, $\Gamma=1.5-2$, the final
emergent nonthermal emission is dominant over the initial injected thermal one, and the
peak energy is $E_{peak}\sim$ a few keV, consistent with observed in GRB 060218. So the
main spectral features of GRB 060218 is reproduced in general by the TRSN shocks. TR bulk
motion is also required to upscatter photons efficiently, and to have low optical depth
at breakout radius, $\tau\sim c/v_{sh}<$ a few, so that photon absorption is unimportant.
The photon arrival times in single-injection cases spread in much shorter periods as
opposed to the observed $10^3$s duration, which is due to the anisotropic geometry. In
our model, the nonthermal emission duration should be comparable to the thermal one.

\section{Low-luminosity GRB: trans-relativistic SN shock breakout} So far Type Ic SNe
are spectroscopically observed to be associated with four GRBs, namely GRBs 980425,
030329, 031203 and 060218. Among them the luminous GRB 030329 appears to be typical,
while the other three show much lower luminosity and lower redshifts ($z<1$). The three
low-luminosity GRBs (LLGRBs) show similar gamma-ray emission properties: they all show
low (isotropic) energy, $E_{\gamma,iso}<10^{50}$erg ($10^{52-54}$erg for typical GRBs);
the light curves are smooth without multiple peaks; the spectra could be fit with power
law with some cutoff at few hundreds keV or lower. These properties are consistent with
nonthermal flashes from TRSN shock breakout discussed above. Furthermore, their
afterglows also show hints for mildly relativistic ejecta. In GRB 980425, the radio
afterglow modelling implies the Lorentz factor and kinetic energy of the SN ejecta as
$\Gamma\approx1.5$ and $E_k\approx5\times10^{49}$, respectively \cite{980425radio}; The
relatively flat light curve of its X-ray afterglow up to $\sim100$ days after the burst
is consistent with a long coasting phase (due to low medium density) of a mildly
relativistic shell with energy of a few $10^{49}$erg \cite{Waxman04}. Similarly, the
X-ray afterglow of GRB 031203 also had an early similar flat light curve \cite{724}.

There are also some diversities in LLGRBs. GRB 980425 shows only short duration,
$\sim30$s, and very low energy, $E_{\gamma,iso}<10^{48}$erg, in contrast with GRB 060218.
GRB 031203 also has duration of tens of seconds. These can be interpreted as that GRBs
980425 and 031203 are produced by relevant TRSNe breakout at the progenitor stellar
surface other than optically thick stellar winds. In this case the pre-shock medium
density and hence the backscattering rate decreases sharply with time. Actually, the
modelling of X-ray/radio afterglow of GRB 980425 indicates only a optically thin wind
\cite{980425radio,Waxman04}.

From the above lines of reasoning, we propose that these three LLGRBs come from photon
acceleration at TRSN shock breakout, which are distinct from the typical GRBs in
mechanism.

\section{Summary}
\begin{itemize}

\item Shock acceleration of thermal photons at TRSN shock breakout can produce
gamma/X-ray flash. This mechanism requires only TR outflow.

\item Both nonthermal and thermal emission (as well as the early, $<1$ day, UV/O emission
and the late, $>10^4$s, power-law-decay X-ray afterglow in GRB 060218/SN 2006aj
\cite{obs,Waxman07}) could be generated in the same context of TRSN shock breakout. This
suggests GRB 060218 as the first detection of the thermal flash from SN shock breakout.

\item LLGRBs could be produced by SN shock breakout with TR ejecta, distinct from typical
GRBs with ultra-relativistic outflows.

\item TRSNe appear to be a new SN category. Large fraction of energy in TR ejecta is
difficult to come from shock acceleration in stellar surface in the case of spherical SN
explosion. TRSNe might always be related to anisotropic hypernovae, but what produces
$\gtrsim10^{49}$erg in TR ejecta is still an open question.

\end{itemize}


\begin{thebibliography}{999}
\bibitem{history} S. A. Colgate, {\em ApJ}, {\bf 187} (1974) 333; R. I. Klein, R. A. Chevalier, {\em ApJ}, {\bf 223} (1978) L109;
L. Ensman, A. Burrows, {\em ApJ}, {\bf 393} (1992) 742.

\bibitem{obs} S. Campana, et al. {\em Nature}, {\bf 442} (2006) 1008.

\bibitem{Waxman07} E. Waxman, P. Meszaros, S. Campana, {\em ApJ} submitted 2007, arXiv:astro-ph/0702450

\bibitem{ph-acc} X.-Y. Wang, Z. Li, E. Waxman, P. Meszaros, {\em ApJ} accepted 2006, arXiv:astro-ph/0608033

\bibitem{980425radio} S. R. Kulkarni, et al. {\em Nature}, {\bf 395} (1998) 663; E. Waxman, A. Loeb, {\em ApJ}, {\bf 515}
(1999) 721; Z.-Y. Li, R.A. Chevalier, {\em ApJ}, {\bf 526} (1999) 716.

\bibitem{Waxman04}
E. Waxman, {\em ApJ}, {\bf 605} (2004) L97.

\bibitem{724}
D. Watson, et al. {\em ApJ}, {\bf 605} (2004) L101.
\end{thebibliography}
\end{document}